\documentclass[10pt
,a4paper
]{article} % KOMA-Script article scrartcl

\usepackage{lipsum}
\usepackage{url}
\usepackage[natbib=true	,style=numeric]{biblatex}
\usepackage{tabularx}
\usepackage{mathtools}
\usepackage{float}
\usepackage{underscore}
\usepackage{caption}
\usepackage{standalone}
\usepackage{rotating}
\usepackage{tabularx}
\usepackage{dirtree}
\newcommand{\tableheadline}[1]{\multicolumn{1}{c}{#1}}

\usepackage{xspace}

\usepackage{chngcntr}
\usepackage{import}

\bibliography{bib} 

\newcommand\wordcount{ \immediate\write18{texcount -sub=section
    \jobname.tex | grep "Section" | sed -e 's/+.*//' | sed -n
    \thesection p > 'count.txt'} \input{count.txt}words}

\usepackage{booktabs}	

\usepackage[left=1.5in,right=1.5in,top=0.5in]{geometry}

\newcommand{\mypddl}{\textsc{myPddl}\xspace}
\newcommand{\mypddlclojure}{\textsc{myPddl-clojure}\xspace}
\newcommand{\mypddlsyntax}{\textsc{myPddl-syntax}\xspace}
\newcommand{\mypddldiagram}{\textsc{myPddl-diagram}\xspace}
\newcommand{\mypddlnew}{\textsc{myPddl-new}\xspace}
\newcommand{\mypddlide}{\textsc{myPddl-ide}\xspace}
\newcommand{\mypddlsnippet}{\textsc{myPddl-snippet}\xspace}
\newcommand{\mypddldistance}{\textsc{myPddl-distance}\xspace}
\newcommand{\pddlstudio}{\textsc{pddl studio}\xspace}
\newcommand{\itsimple}{\textsc{itSimple}\xspace}
\newcommand{\pddlmode}{\textsc{pddl}-mode\xspace}
\newcommand{\pddl}{\textsc{pddl}\xspace}
\newcommand{\uml}{\textsc{uml}\xspace}
\newcommand{\ide}{\textsc{ide}\xspace}
\newcommand{\sublimetext}{Sublime Text\xspace}

\newcommand\blfootnote[1]{%
  \begingroup
  \renewcommand\thefootnote{}\footnote{#1}%
  \addtocounter{footnote}{-1}%
  \endgroup
}

\begin{document}

\title{\normalfont Planning in the Wild: Modeling Tools for PDDL}
\author{
  Volker Strobel\thanks{Contact email address: volker.strobel87@gmail.com}
\and Alexandra Kirsch\thanks{University of T\"ubingen}}
\date{}
\maketitle

\blfootnote{The final publication is available at Springer via http://dx.doi.org/10.1007/978-3-319-11206-0_27}

Writing and maintaining planning problems, specified in the widely
used \emph{Planning Domain Definition Language} (\pddl) can be
difficult, time-consuming, and error-prone. One reason seems to be the
missing support by engineering tools. The present study proposes
\mypddl~-- a modular toolkit for developing and manipulating \pddl
domains and problems. To evaluate \mypddl, we compare it to existing
knowledge engineering tools for \pddl and experimentally assess its
usefulness for novice \pddl users.

\section{Introduction}

A large community of researchers dedicate their efforts to Artificial
Intelligence (AI) planning. However, the process made in this
community is often ignored when it comes to developing complete AI
systems. Planning is a fundamental cognitive function that is useful
for most systems claiming to be intelligent, such as autonomous robots
or decision support systems. This raises the question why planning is
not used in more systems. We believe that one reason is the gap
between modeling textbook toy problems and modeling complex,
real-world problems.

The standard AI planning language \pddl (\emph{Planning Domain
  Definition Language}) differentiates between domain files with
definitions of types, predicates and actions, and problem files with
definitions of objects and goals. Realistic scenarios contain hundreds
of objects, different agents with different capabilities, able to
perform a large variety of actions. Modeling such worlds soon gets
confusing.

This problem is not specific to planning, but poses a challenge to
software engineering in general. As projects grow in size, developers
have to be supported with appropriate tools to keep track of the
overall structure.

This paper proposes \mypddl, a set of tools for modeling large \pddl
domains and associated problems. Section~2 discusses existing tools
for \pddl. Section~3 presents \mypddl 's modules and design
principles. They are evaluated with a user test in
Section~4. Section~5 concludes with an outlook on further steps
necessary to improve the availability of planning for intelligent
system development.

\section{Related Work}

There have been some attempts to provide modeling tools for
\pddl. This section introduces the three most sophisticated
tools we found.

\pddlstudio \cite{plch2012inspect} is an application for creating and
managing \pddl projects, i.e. a collection of \pddl
files. \pddlstudio 's integrated development environment (\ide) was
inspired by Microsoft Visual Studio and imperative programming
paradigms. Its main features are syntax highlighting, error detection,
context sensitive code completion, code folding, project management,
and planner integration. \pddlstudio 's error detection can recognize
both syntactic (missing keywords, parentheses, etc.) and semantic
(wrong type of predicate parameters, misspelled predicates, etc.)
errors.

A major drawback of \pddlstudio is that it is not updated regularly
and only supports \pddl 1.2. Later \pddl versions contain several
additional features such as durative actions, numeric fluents, and
plan metrics \cite{edelkamp2004pddl2}.

\itsimple \cite{vaquero2005itsimple} follows a graphical approach
using Unified Modeling Language (\uml) diagrams. In the process
leading up to \itsimple, \textsc{uml.p} (\uml in a Planning Approach)
was proposed, a \uml variant specifically designed for modeling
planning domains and problems \cite{vaquero2006use}.

\itsimple's modeling workflow is unidirectional as changes in the
\pddl domain do not affect the \uml model and \uml models have to be
modeled manually, meaning that they cannot by generated from
\pddl. However, \textcite{tonidandel2006reading} present a translation
process from a \pddl domain specification to an object-oriented
\textsc{uml.p} model as a possible integration for \itsimple. This
translation process makes extensive semantic assumptions for \pddl
descriptions. For example, the first parameter in the
\texttt{:parameters} section of an action is automatically declared as
a subclass of the default class \texttt{Agent}, and the method is
limited to predicates with a maximum arity of two. The currently
version of \itsimple does not include the translation process from
\pddl to \uml.

Starting in version 4.0, \itsimple expanded its features to allow the
creation of \pddl projects from scratch (i.e. without the \uml to
\pddl translation process) \cite{vaquero2012itsimple4}. Thus far, the
\pddl editing features are basic. A minimal syntax highlighting
feature recognizes \pddl keywords, variables, and comments. \itsimple
also provides templates for \pddl constructs, such as requirement
specifications, predicates, actions, initial state, and goal
definitions.

Both \pddlstudio and \itsimple do not build on existing editors and
therefore cannot fall back on refined implementations of features that
have been modified and improved many times throughout their existence.

\pddlmode\footnote{http://rakaposhi.eas.asu.edu/planning-
  list-mailarchive/msg00085.html } for the widely used Emacs editor
builds on the sophisticated features of Emacs and uses its
extensibility and customizability. It provides syntax highlighting by
way of basic pattern matching of keywords, variables, and
comments. Additional features are automatic indentation and code
completion as well as bracket matching. Code snippets for the creation
of domains, problems, and actions are also available. Finally,
\textsc{pddl}-mode keeps track of action and problem declarations by
adding them to a menu and thus intending to allow for easy and fast
code navigation.

\pddlmode for Emacs supports \textsc{pddl} versions up to
2.2, which includes derived predicates and timed initial predicates
\cite{edelkamp2004pddl2}, but does not recognize later features like
object-fluents.

In sum, there is currently no tool available supporting all features
of \pddl 3.1, nor all the steps in the modeling process.

\section{myPDDL}

\mypddl is designed as a modular framework. We first introduce the
implemented modules and then explain their details with respect to
design guidelines for knowledge engineering tools.

\subsection{Modules}

\begin{description}
\item[\mypddlide] is an integrated development environment for the use
  of \mypddl in the text and code editor \emph{Sublime
    Text}\footnote{\url{http://www.sublimetext.com}}. Since
  \mypddlsnippet and \textsc{-syntax} are devised explicitly for
  \sublimetext, their integration is implicit. The other tools can be
  used independently of \sublimetext with the command-line interface
  and any \pddl file, but were also integrated into the editor.

\item[\mypddlsyntax] is a context-aware syntax highlighting feature
  for \sublimetext. It distinguishes all \pddl constructs up to
  version 3.1. Using regular expressions that can recognize both the
  start and the end of code blocks by means of a sophisticated pattern
  matching heuristic, \mypddlsyntax identifies \pddl code blocks and
  constructs and divides them into so called scopes, i.e. named
  regions. \sublimetext colorizes the code elements via the assigned
  scope names and in accordance with the current color scheme. These
  scopes allow for a fragmentation of the \pddl files, so that
  constructs are only highlighted if they appear in the correct
  context. Thus missing brackets, misplaced expressions and misspelled
  keywords are visually distinct and can be identified (see
  Figure~\ref{fig:syntax}).

  \begin{figure}
    \centering
    \includegraphics[width=0.6\textwidth]{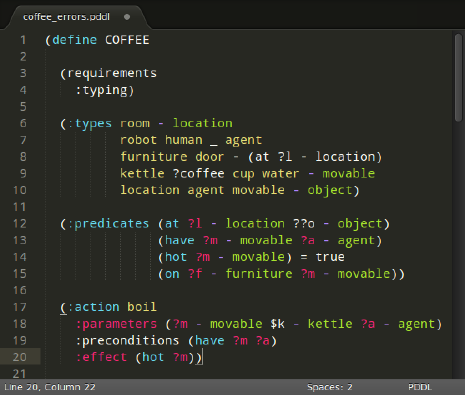}
    \caption{Syntax highlighting using \mypddlide. White
      text contains errors.}
\label{fig:syntax}
  \end{figure}

\item[\mypddlnew] helps to organize \pddl projects by generating the
following folder structure:

\begin{figure}[h] 
  \dirtree{%
  .1 project-name/.
  .2 domains/.
  .2 problems/.
  .3 p01.pddl.
  .2 solutions/.
  .2 domain.pddl.
  .2 README.md.
  }
\end{figure}

The domain file \texttt{domain.pddl} and the problem file
\texttt{p01.pddl} initially contain corresponding \pddl skeletons
which can also be customized. All problem files that are associated
with one domain file are collected in the folder
\texttt{problems/}. \texttt{README.md} is a Markdown file, which is
intended for (but not limited to) information about the author(s) of
the project, contact information, informal domain and problem
specifications, and licensing information.  Markdown files can be
converted to \textsc{html} by various hosting services (like GitHub or
Bitbucket).

\item[\mypddlsnippet] provides code skeletons, i.e. templates for
  often used pddl constructs such as domains, problems, type and
  function declarations, and actions. They can be inserted by typing a
  triggering keyword.

\item[\mypddlclojure] provides a preprocessor for \pddl files to
  bypass \pddl's limited mathematical capabilities, thus reducing
  modeling time without overcharging planning algorithms. We decided
  to use Clojure \cite{hickey2008clojure}, a modern Lisp dialect that
  runs on the Java Virtual Machine (\textsc{jvm})
  \cite{lindholm2014java}, facilitating input and output of the
  Lisp-style \pddl constructs.

\item[\mypddldistance] provides special preprocessing functions for
  distance calculations. For domains with spatial components, the
  distance of objects is often important and should not be omitted in
  the domain model. However, calculating distances from coordinates
  requires the square root function, which is not supported by \pddl
  (it only supports the four basic arithmetic operators). More
  sophisticated calculations can be achieved with the supported
  operators, but the solutions are rather inefficent and inelegant
  \cite{parkinson2012increasing}. By calculating the distances offline
  and including them as additional predicates in the problem file
  using \mypddldistance, the distances between objects are given to
  the planner as part of the problem description.

\item[\mypddldiagram] generates a \textsc{png} image from a \pddl
  domain file as shown in Figure~\ref{fig:diagram}. The diagrammatic
  representation of textual information helps to quickly understand
  the connection of hierarchically structured items and should thus be
  able to simplify the communication and collaboration between
  developers. In the process of generating the diagrams, a copy of the
  \pddl file is created, so that a simple version control is also
  included.

  \begin{figure}
    \centering
    \includegraphics[width=1\textwidth]{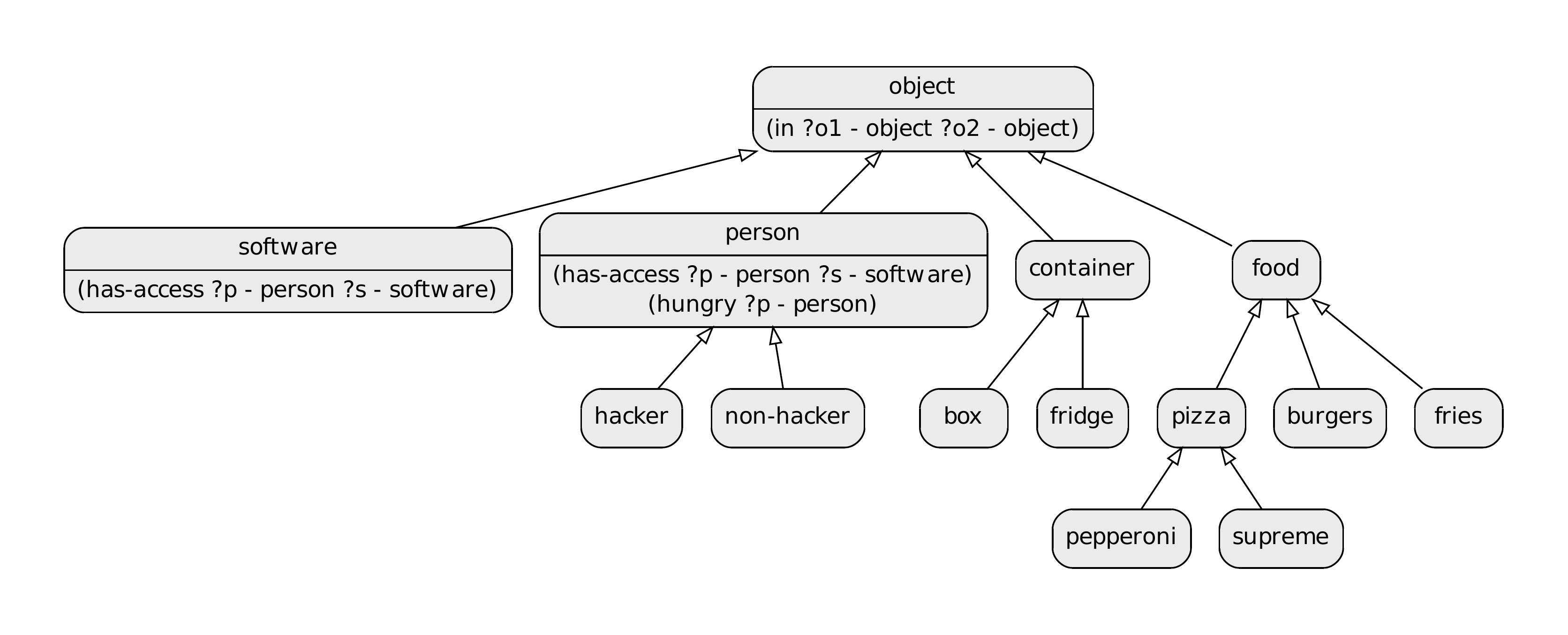}
    \caption{Type diagram generated by \mypddldiagram}
    \label{fig:diagram}
  \end{figure}

\end{description}

\subsection{Design Principles}

As guidelines for design decisions, we used the seven criteria for
knowledge engineering tools proposed by Shah et
al. \cite{shah2013knowledge} as well as general usability
principles. Operationality instantiates, whether the generated models
can improve the planning performance. This is not a design principle
for \mypddl, because we assume that \mypddl does not improve the
quality (with respect to planning performance) of the resulting \pddl
specifications. Therefore, we replaced this criterion with functional
suitability from the \textsc{iso/iec} 25010 standard, which is defined
as “the degree to which the software product provides an appropriate
set of functions for specified tasks and user objectives”
(\textsc{iso} 25010 6.1.1). \mypddl supports the current version 3.1
of \pddl. It encompasses and exceeds most of the functionality of the
existing tools. It specifically supports basic editor features with a
high customizability as well as visualization support.  Collaboration:
With the growing importance of team work and team members not
necessarily working in the same building, or in the same country,
there is an increasing need for tools supporting the collaboration
effort. In developing \mypddl, this need was sought to be met by
\mypddldiagram. Complex type hierarchies can be hard to overlook,
especially if they were constructed by someone else. Therefore, a good
way of tackling this problem seemed to be by providing a means to
visualize such hierarchies in the form of type diagrams.

\begin{description}

\item[Experience]: \mypddl was designed specifically for users with a
  background in AI, but not necessarily in \pddl. The tools are
  similar to standard software engineering tools and should thus be
  easily learnable. The user evaluation (Section 4.2) confirms that
  \mypddl helps novices in \pddl to master planning task modeling. In
  addition, it is also possible to customize \mypddl so as to adapt
  its look and feel to other programs one is already familiar with, or
  simply to make it more enjoyable to use. The project
  site\footnote{\url{http://pold87.github.io/myPDDL}} provides \mypddl
  video introductions and a manual to get started quickly.

\item[Efficiency]: All \mypddl tools are intended to increase the
  efficiency with which \pddl files are created. \mypddlsnippet
  enables the fast creation of large and correct code skeletons that
  only need to be complemented. \mypddlsyntax can reduce the time
  spent on searching errors. Code folding allows users to hide
  currently irrelevant parts of the code and automatic indentation
  increases its readability. To easily keep track of all the parts of
  a project, folders are automatically created and named with
  \mypddlnew. \mypddlclojure and \textsc{-distance} allow
  for a straightforward inclusion of numerical values in the problem
  definition.

\item[Debugging]: \mypddl\textsc{-syntax} highlights all syntactically
  correct constructs and leaves all syntactical errors
  non-highlighted. In contrast, \pddlmode for Emacs and itSimple only
  provide basic syntax highlighting for emphasizing the
  structure. \pddlstudio explicitly detects errors, but the user is
  immediately prompted when an error is detected. Often, such error
  messages are premature, for example, just because the closing
  parenthesis was not typed yet, does not mean it was
  forgotten. \mypddl indicates errors in a more subtle way: syntactic
  errors are simply not highlighted, while all correct \pddl code
  is. The colors are customizable, so that users can choose how
  prominent the highlighting sticks out.

\item[Maintenance]: The possibility to maintain \pddl files is a key
  aspect of \mypddl.  The automatically generated type diagram
  (\mypddldiagram) gives an overview of the domain structure and
  thereby serves as a continuous means of documentation. Helping to
  understand foreign code, though, it follows logically that
  \mypddldiagram also helps in coming back and changing one’s own
  models if some time has elapsed since they were last edited. The
  basic revision control feature of \mypddldiagram keeps track of
  changes, making it easy to revert to a previous domain
  version. Furthermore, \mypddlnew encourages adhering to an organized
  project structure and stores corresponding files at the same
  location.  The automatically created readme file can induce the user
  to provide further information and documentation about the \pddl
  project.  Support: \mypddlide can be installed using \sublimetext's
  Package
  Control\footnote{\url{https://sublime.wbond.net/about}}. This allows
  for an easy installation and staying up-to-date with future
  versions. In order to provide global access and with it the
  possibility for developing an active community, the project source
  code is hosted on
  GitHub\footnote{https://github.com/Pold87/myPDDL}. Additionally, the
  project site provides room for discussing features and reporting
  bugs.

\end{description}
\section{Validation and Evaluation}

To assess the utility of \mypddl, we used the criteria listed in
Section 3.2. The functional suitability was evaluated using a
benchmark validation, comparing \mypddl’s functionality with the tools
described in Section 2. The criteria collaboration, experience,
efficiency, and debugging were evaluated in a user test. The \mypddl
components supporting maintenance are the same ones that are used in
the user test, but their long-term usage is difficult to evaluate. The
support criterion depends primarily on the infrastructure, which has
been established as explained in 3.2.

\subsection{Benchmark Validation}

Functional suitability encompasses the set of functions to meet the
user objectives. The tools of Section 2 basically all follow the same
objectives as \mypddl: creating \pddl domains and problems. The
features offered by each tool are summarized in Table
\ref{tool-comp}. Besides supporting the latest \pddl version, a
strength of \mypddl is its high customizability, which comes with the
\sublimetext editor. Being the only one of the four tools capable of
visualizing parts of the \pddl code, it must be understood as
complementary to \itsimple, which takes the opposite approach of
transforming \uml diagrams into \pddl files. The fact that \mypddl
does not check for semantic errors is not actually a drawback as
planners will usually detect semantic errors. All in all, \mypddl
combines the most useful tools of \pddlstudio, \itsimple, and
\pddlmode for Emacs and strives to support the planning task engineer
during all phases of the modeling process. Additionally, it features
some unique tools, such as domain visualization and an interface with
a programming language. It can therefore be concluded that \mypddl
provides an appropriate set of functions for developing \pddl files
and is thus functionally suitable.

\begin{sidewaystable}[htb]
\centering
\footnotesize
\begin{tabularx}{\textwidth}{lX|llll}
  \tableheadline{Feature}             & \tableheadline{Function}                              & \tableheadline{\pddlstudio} & \tableheadline{\itsimple} & \tableheadline{\pddlmode} & \tableheadline{\mypddl} \\
  \hline
  latest supported \pddl version      & considering recent \pddl features                     & 1.2                         & 3.1                       & 2.2                       & 3.1                     \\
  syntax highlighting                 & supporting error detection and code navigation        & yes                         & basic                     & basic                     & yes                     \\
  semantic error detection            & supporting error detection                            & yes                         & no                        & no                        & no                      \\
  automatic indentation               & supporting readability and navigation                 & no                          & no                        & yes                       & yes                     \\
  code completion                     & speeding-up the knowledge engineering process         & yes                         & no                        & yes                       & yes                     \\
  code snippets                       & speeding-up the knowledge engineering process         & no                          & yes                       & yes                       & yes                     \\
                                      & externalizing user's memory                           &                             &                           &                           &                         \\
  code folding                        & supporting keeping an overview of the code structure  & yes                         & no                        & yes                       & yes                     \\
  domain visualization                & supporting fast understanding of the domain structure & no                          & planned                        & no                        & yes                     \\
  project management                  & supporting keeping an overview of associated files    & yes                         & yes                       & no                        & yes                     \\
  \uml to \pddl code translation      & supporting initial modeling                           & no                          & yes                       & no                        & yes                     \\
  planner integration                 & allowing for convenient planner access                & basic                       & yes                       & no                        & basic                    \\
  plan visualization                  & supporting understanding and crosschecking the plan   & no                          & yes                       & no                        & no                      \\
  dynamic analysis                  & supporting dynamic domain analysis                    & no                          & yes                       & no                        & no                      \\
  declaration menu                    & supporting code navigation                            & no                          & no                        & yes                       & no                      \\
  interface with programming language & automating tasks                                      & no                          & no                        & no                        & yes                     \\
                                      & extending \pddl's modeling capabilities               &                             &                           &                           &                         \\
  customization features              & acknowledging individual needs and preferences        & basic                       & no                        & yes                       & yes                     \\
\end{tabularx}\caption[Comparison of knowledge engineering
tools]{\label{tool-comp}Comparison of knowledge engineering tools and their features.}

\end{sidewaystable}

\subsection{User Evaluation}

The two most central modules of \mypddl are \mypddlsyntax and
\mypddldiagram, since they support collaboration, efficiency, and
debugging independently of the user’s experience with \pddl. To
evaluate their usability, they will be evaluated in a user study.

\subsubsection{Procedure}

We invited eight participants\footnote{In Usability Engineering, a
  typical number of participants for user tests is five to
  ten. Studies have shown that even such small sample sizes identify
  about 80\,\% of the usability problems
  \cite{nielsen1994estimating,hwang2010number}. Our study design
  required at least eight participants.} to a user test (three female,
average age 22.9, standard deviation 0.6), who had some basic
experience with at least one Lisp dialect (in order not to be confused
with the many parentheses), but no experience with \pddl or AI
planning in general.

No earlier than 24 hours before the experiment was to take place,
participants received the web link to a 30-minute interactive video
tutorial on AI planning and \pddl. This method was chosen in
order not to pressure the participant with the presence of an
experimenter when trying to understand the material.

We defined four tasks: two debugging tasks and two type hierarchy
tasks asking for details of a given domain (e.g. ``Can a Spleus be
married to a Schlok?'').  As a within subjects design was considered
most suited (to control for individual differences within such a small
sample), it was necessary to construct two tasks (matched in
difficulty) for each of these two types to compare the effects of
having the tools available. The two tasks to test syntax highlighting
presented the user with domains that were 54 lines in length,
consisted of 1605 characters and contained 17 errors each. Errors were
distributed evenly throughout the domains and were categorized into
different types. The occurrence frequencies of these types were
matched across domains as well, to ensure equal difficulty for both
domains. To test the type diagram generator, two fictional domains
with equally complex type hierarchies consisting of non-words were
designed (five and six layers in depth, 20 and 21 types). The domains
were also matched in length and overall complexity (five and six
predicates with approximately the same distribution of arities, one
action with four predicates in the precondition and two and three
predicates in the effect).

Each participant started either with a debugging or type hierarchy
task and was given the \mypddl tools either in the first two tasks or
the second two tasks, so that each participant completed each task
type once with and once without \mypddl. This results in $2$ (first
task is debugging or hierarchy) $\times$ $2$ (task variations for
debugging and hierarchy) $\times$ $2$ (starting with or without \mypddl)
$= 8$ individual task orders, one per participant.

For the debugging tasks, participants were given six
minutes\footnote{A reasonable time frame tested on two pilot tests.}
to detect as many of the errors as possible. They were asked to record
each error in a table (pen and paper) with the line number and a short
comment and to immediately correct the errors in the code if they knew
how to, but not to dwell on the correction otherwise. For the type
hierarchy task, participants were asked to answer five questions
concerning the domains, all of which could be facilitated with the
type diagram generator, but one of which also required looking into
the code. Participants were told that they should not feel pressured
to answer quickly, but to not waste time either. Also they were asked
to say their answer out loud as soon as it became evident to
them. They were not told that the time it took them to come up with an
answer was recorded, since this could have made them feel pressured
and thus led to more false answers. At the end of the usability test
they were asked to evaluate the perceived usability of \mypddl using
the system usability scale \cite{brooke1996sus}.

\subsubsection{Results}

\begin{itemize}
\item Debugging Tasks

  As shown in Figure 3, on average participants found 7.6 errors
  without syntax highlighting and 10.3 errors with syntax highlighting
  (i.e. approximately 36\,\% more errors were found with syntax
  highlighting). Two participants remarked that the syntax
  highlighting colors confused them and that they found them more
  distracting than helpful. One of them mentioned that the contrast of
  the colors used was so low that they were hard for her to
  distinguish. She found the same number of errors with and without
  syntax highlighting. The other of the two was the only participant
  who found less errors with syntax highlighting than without it. With
  \mypddlsyntax, two participants found all errors in the
  domain, while none achieved this without syntax highlighting.

\begin{figure}[h]
  \centering
  \hspace{1.7cm}
  \includegraphics[width=0.6\textwidth]{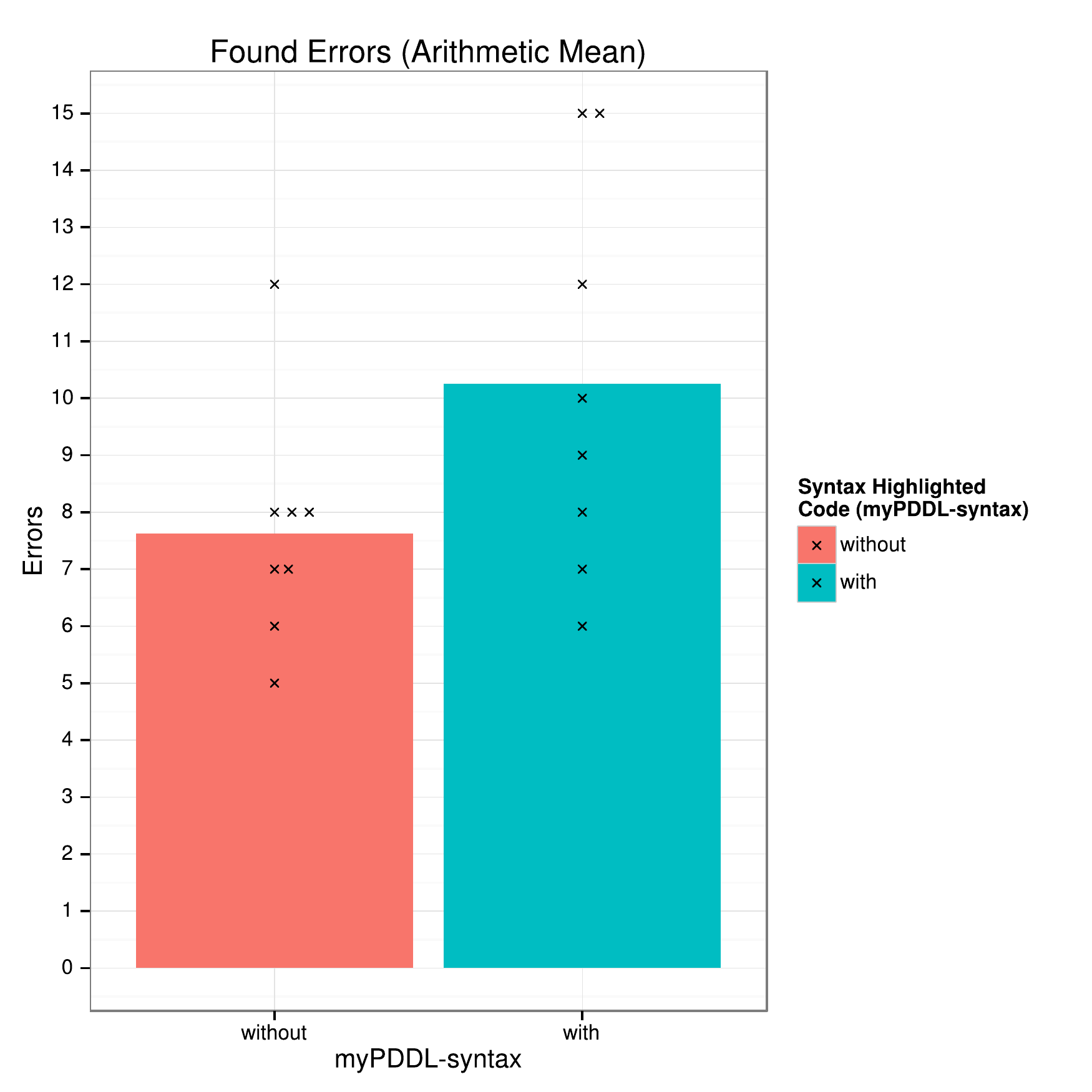}
  \caption[Diagram of detected errors]{Comparison of detected errors with and without the syntax
    highlighting feature. Each cross ($\times$) shows the data value
    of one participant. The bars display the arithmetic mean.}
\label{fig:found-errors-combined}
\end{figure}

\item Type Hierarchy Tasks

  Figure~\ref{fig:task-completions-agg} shows the geometric
  mean\footnote{The geometric mean is a more accurate measure of the
    mean for small sample sizes as task times have a strong tendency
    to be positively skewed \cite{sauro2012quantifying}.} of the
  completion time of successful tasks for each question with and
  without the type diagram generator.  With the type diagram generator
  participants answered all questions (except Question 4) on average
  nearly twice as fast. The fact that the availability of tools did
  not have a positive effect on task completion times for Question 4
  can probably be attributed to the complexity of this question. In
  contrast to the other four questions, to answer Question 4
  correctly, the participants were required to look at the actions in
  the domain file in addition to the type diagram. Most participants
  were confused by this, because they had assumed that once having the
  type diagram available, it alone would suffice to answer all
  questions. This initial confusion cost some time, thus negatively
  influencing the time on the task.

\begin{figure}[h]
  \centering
  \hspace{1.7cm}
  \includegraphics[width=0.74\textwidth]{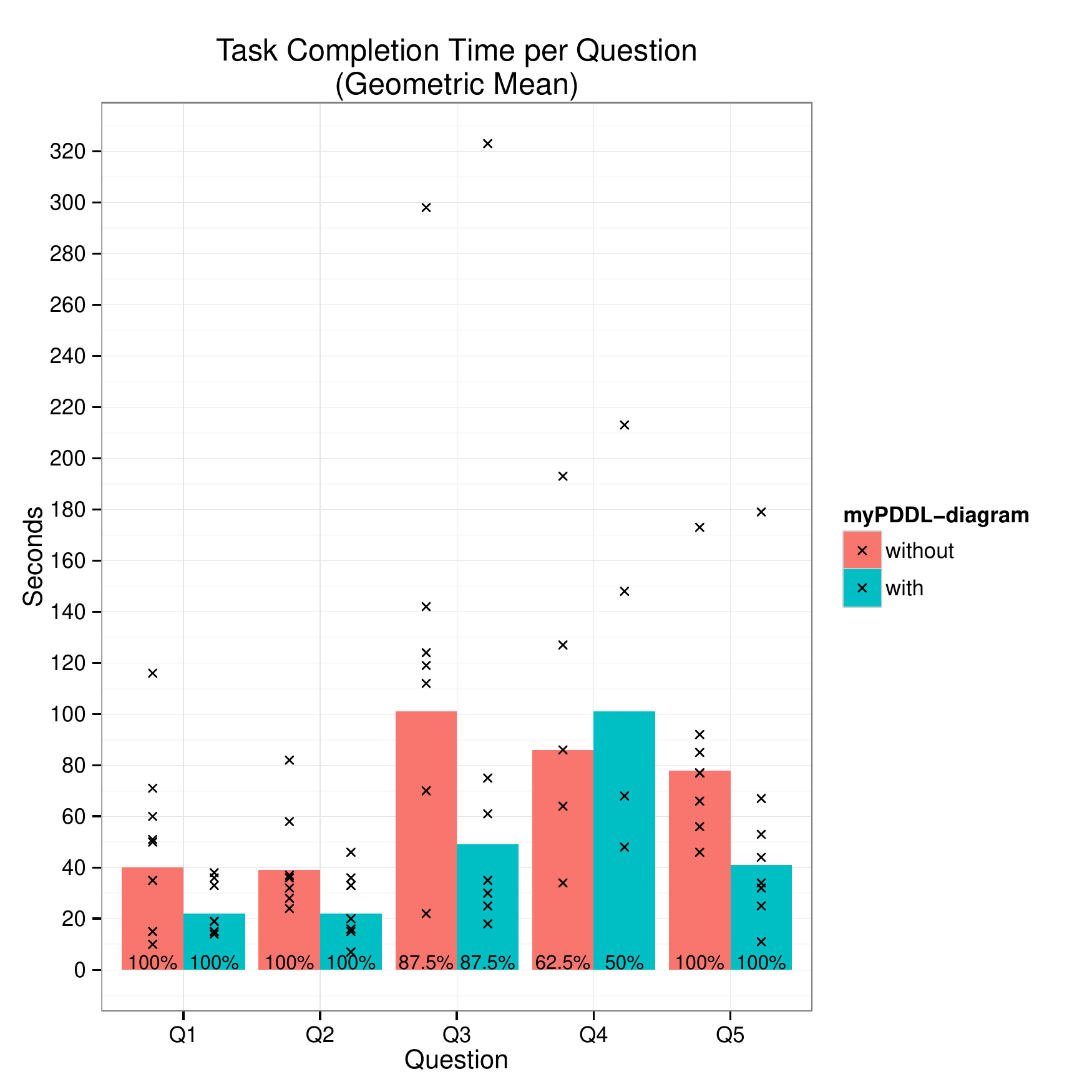}
  \caption[Diagram of the task completion time]{Task completion time for the type hierarchy
tasks. The bars display the geometric mean. The
percent values at the bottom of the bars show
the the percentage of users that completed the
task successfully.}
  \label{fig:task-completions-agg}
\end{figure}

\item System Usability Scale

  \mypddl reached a score of 89.6 on the system usability scale 10 ,
  with a standard deviation of 3.9. Since the overall mean score of
  the system usability scale has an approximate value of 68 with a
  standard deviation of 12.5 \cite{sauro2011practical}, the score of
  \mypddl is well above average with a small standard deviation. A
  score of 89.6 is usually attributed to superior products
  \cite{bangor2008empirical}. Furthermore, 89.6 corresponds
  approximately to a percentile rank of 99.8\,\%, meaning that it has
  a better perceived ease-of-use than 99.8\,\% of the products in the
  database used by Sauro \cite{sauro2011practical}.

\end{itemize}

\section{Conclusion}

\mypddl was designed with the goal to support plan engineers in
modeling domains and planning problems as well as in understanding,
modifying, extending, and using existing planning domains. This was
realized with a set of tools comprising code editing features, namely
syntax highlighting and code snippets, a type diagram generator, and a
distance calculator. To also have all tools accessible from one place,
they were made available in the \sublimetext editor. The different
needs and requirements of knowledge engineers are met by the modular,
extensible, and customizable architecture of the toolkit and
\sublimetext. The evaluation of \mypddl has shown evidence that it
allows a faster understanding of the domain structure, which could be
beneficial for the maintenance and application of existing task
specifications and for the communication between engineers. Users
perceive it as easy and enjoyable to use, and the increase in their
performance when using \mypddl underpins their subjective impressions.
Despite \mypddl already providing a rich modeling environment, there
are still numerous features that could be added in the
future. Especially \mypddlclojure offers multiple interesting further
research directions: It provides a basis for dynamic planning
scenarios. Applications could be the modeling of learning and
forgetting (by adding facts to or retracting facts from a \pddl file)
or the modeling of an ever changing real world via dynamic predicate
lists.  Another way of putting the interface to use would be by making
the planning process more interactive, allowing for the online
interception of planning software in order to account for the needs
and wishes of the end user.

%\section{References}
\newpage
\printbibliography

\end{document}